\documentclass [sigconf]{acmart}
\AtBeginDocument{%
  \providecommand\BibTeX{{%
    \normalfont B\kern-0.5em{\scshape i\kern-0.25em b}\kern-0.8em\TeX}}}

\usepackage{times}
\usepackage{soul}
\usepackage{url}
\usepackage[utf8]{inputenc}
\usepackage{caption}
\usepackage{graphicx}
\usepackage{amsmath}
\usepackage{amsthm}
\usepackage{booktabs}
\usepackage{algorithm}
\usepackage{algorithmic}
\usepackage[switch]{lineno}
\usepackage{multirow}
\usepackage{url}
\usepackage{titlesec}
\usepackage{amsfonts}
\usepackage{nicefrac}
\usepackage{microtype}
\usepackage{xcolor}
\usepackage{amsmath}
\usepackage{graphicx}
\usepackage{enumitem}
\usepackage{multirow}
\usepackage{subcaption}
\usepackage{listings}
\usepackage{tcolorbox}
\usepackage{xspace}
\usepackage{makecell}
\setstcolor{red}
\usepackage{tikz}
\usepackage{longtable}
\usepackage{tablefootnote}
\usepackage[edges]{forest}
\definecolor{hidden-draw}{RGB}{20,68,106}
\definecolor{hidden-pink}{RGB}{255,245,247}
\definecolor{red}{RGB}{255,0,0}

\makeatletter
\DeclareRobustCommand\onedot{\futurelet\@let@token\@onedot}
\def\@onedot{\ifx\@let@token.\else.\null\fi\xspace}

\def\etal{\emph{et al}\onedot}
\makeatother
\usepackage[draft,textsize=footnotesize,textwidth=15mm]{todonotes}

\definecolor{paired-light-blue}{RGB}{198, 219, 239}
\definecolor{paired-dark-blue}{RGB}{49, 130, 188}
\definecolor{paired-light-orange}{RGB}{251, 208, 162}
\definecolor{paired-dark-orange}{RGB}{230, 85, 12}
\definecolor{paired-light-green}{RGB}{199, 233, 193}
\definecolor{paired-dark-green}{RGB}{49, 163, 83}
\definecolor{paired-light-purple}{RGB}{218, 218, 235}
\definecolor{paired-dark-purple}{RGB}{117, 107, 176}
\definecolor{paired-light-gray}{RGB}{217, 217, 217}
\definecolor{paired-dark-gray}{RGB}{99, 99, 99}
\definecolor{paired-light-pink}{RGB}{222, 158, 214}
\definecolor{paired-dark-pink}{RGB}{123, 65, 115}
\definecolor{paired-light-red}{RGB}{231, 150, 156}
\definecolor{paired-dark-red}{RGB}{131, 60, 56}
\definecolor{paired-light-yellow}{RGB}{231, 204, 149}
\definecolor{paired-dark-yellow}{RGB}{141, 109, 49}

\definecolor{bg1}{HTML}{FF9966}
\definecolor{bg2}{HTML}{CCE5FF}
\definecolor{bg3}{HTML}{FFCC99}
\definecolor{bg4}{HTML}{FFC107}
\definecolor{bg5}{HTML}{FFCCCC}
\definecolor{bg6}{HTML}{D5E8D4}
\definecolor{bg7}{HTML}{eeeeee}
\definecolor{bg8}{HTML}{cdeb8b}
\definecolor{bg9}{HTML}{dae8fc}
\definecolor{bg10}{HTML}{a2e6eb}

\definecolor{bg31}{HTML}{FFCDD2} 

\definecolor{bg32}{HTML}{F8BBD0}

\definecolor{bg33}{HTML}{E1BEE7} 

\definecolor{bg34}{HTML}{D7CCC8} 

\definecolor{bg35}{HTML}{B2DFDB} 

\definecolor{bg36}{HTML}{A5D6A7} 

\definecolor{bg37}{HTML}{FFF9C4} 

\definecolor{bg38}{HTML}{FFECB3} 

\definecolor{bg111}{HTML}{CB6843}

\definecolor{bg112}{HTML}{D77C5C}

\definecolor{bg113}{HTML}{E28E6E}
\definecolor{bg114}{HTML}{E89F7D}
\definecolor{bg115}{HTML}{EDAE8A}
\definecolor{bg116}{HTML}{F0BA95}
\definecolor{bg117}{HTML}{F3C29F}
\definecolor{bg118}{HTML}{F6CCAA}
\definecolor{bg119}{HTML}{F8D5B3}
\definecolor{bg120}{HTML}{FADCBD}
\definecolor{bg121}{HTML}{FCE6C7}

\definecolor{bg39}{HTML}{FFE0B2} 

\definecolor{bg40}{HTML}{3CB371} 

\definecolor{bg43}{HTML}{ffe5d9}

\definecolor{bg15}{HTML}{7FFFD4}

\definecolor{bg17}{HTML}{F0FFFF}

\definecolor{bg18}{HTML}{F5FFFA}

\definecolor{bg19}{HTML}{F8F8FF}

\definecolor{bg20}{HTML}{FFFFFF}

\definecolor{bg21}{HTML}{E1F5FE}

\definecolor{bg22}{HTML}{B3E5FC}

\definecolor{bg23}{HTML}{81D4FA}

\definecolor{bg24}{HTML}{4FC3F7}

\definecolor{bg25}{HTML}{29B6F6}

\definecolor{bg26}{HTML}{03A9F4}

\definecolor{bg27}{HTML}{039BE5}

\definecolor{bg28}{HTML}{0288D1}

\definecolor{bg29}{HTML}{0277BD}

\definecolor{bg30}{HTML}{01579B}

\definecolor{bg16}{HTML}{FFCC99}

\definecolor{pg51}{HTML}{E8F5E9} 
\definecolor{pg52}{HTML}{C8E6C9} 
\definecolor{pg53}{HTML}{B9F6CA} 
\definecolor{pg54}{HTML}{A9DFBF} 
\definecolor{pg55}{HTML}{BCF5A6} 

\definecolor{pg56}{HTML}{BEF1CE} 
\definecolor{pg57}{HTML}{CEF6EC} 
\definecolor{pg58}{HTML}{B7F0B1} 
\definecolor{pg59}{HTML}{B1F2B5} 
\definecolor{pg60}{HTML}{9DF3C4} 

\definecolor{pg61}{HTML}{DEF7E0} 
\definecolor{pg62}{HTML}{E8F8DC} 

\definecolor{pg63}{HTML}{EBF7E7} 
\definecolor{pg64}{HTML}{F0FDF4} 

\definecolor{pg65}{HTML}{F1FEE7} 
\definecolor{pg66}{HTML}{F7FFF6} 
\definecolor{pg67}{HTML}{FCFFE7} 
\definecolor{pg68}{HTML}{F4FFD2} 
\definecolor{pg69}{HTML}{EEFFE2} 
\definecolor{pg70}{HTML}{E3FDF5} 

\definecolor{connect-color}{RGB}{0,0,0}
\definecolor{middle-color}{RGB}{255,255,255}
\definecolor{leaf-color}{RGB}{173,216,230}
\definecolor{line-color}{RGB}{25,25,112}

\begin{document}

\title{Exploring the Impact of Large Language Models on Recommender Systems: An Extensive Review}

\author{Arpita Vats}
\affiliation{%
  \institution{Santa Clara University}
  \streetaddress{Santa Clara}
  \city{Santa Clara}
  \country{USA}}

\author{Vinija Jain\textsuperscript{*}}
\affiliation{%
  \institution{Stanford University\\Amazon}
  \city{Palo Alto}
  \country{USA}
  \email{}
}

\author{Rahul Raja}
\affiliation{%
 \institution{Carnegie Mellon University}
 \city{Pittsburgh}
 \country{USA}
 \email{}
}
\author{Aman Chadha\textsuperscript{*}}\thanks{\textsuperscript{*}Work does not relate to position at Amazon.}
\affiliation{%
\institution{Stanford University\\Amazon GenAI}
  \city{Palo Alto}
  \country{USA}
  \email{}
  }


\begin{abstract}
The paper underscores the significance of Large Language Models (LLMs) in reshaping recommender systems, attributing their value to unique reasoning abilities absent in traditional recommenders. Unlike conventional systems lacking direct user interaction data, LLMs exhibit exceptional proficiency in recommending items, showcasing their adeptness in comprehending intricacies of language. This marks a fundamental paradigm shift in the realm of recommendations. Amidst the dynamic research landscape, researchers actively harness the language comprehension and generation capabilities of LLMs to redefine the foundations of recommendation tasks. The investigation thoroughly explores the inherent strengths of LLMs within recommendation frameworks, encompassing nuanced contextual comprehension, seamless transitions across diverse domains, adoption of unified approaches, holistic learning strategies leveraging shared data reservoirs, transparent decision-making, and iterative improvements. Despite their transformative potential, challenges persist, including sensitivity to input prompts, occasional misinterpretations, and unforeseen recommendations, necessitating continuous refinement and evolution in LLM-driven recommender systems.
\end{abstract}
\maketitle

\section{Introduction}
Recommendation systems are crucial for personalized content discovery, and the integration of Large Language Models (LLMs) in Natural Language Processing (NLP) is revolutionizing these systems~\cite{huang2023recommender}. LLMs, with their language comprehension capabilities, recommend items based on contextual understanding, eliminating the need for explicit behavioral data. The current research landscape is witnessing a surge in efforts to leverage LLMs for refining recommender systems, transforming recommendation tasks into exercises in language understanding and generation. These models excel in contextual understanding, adapting to zero and few-shot domains~\cite{cui2022m6rec}, streamlining processes, and reducing environmental impact. This paper delves into how LLMs enhance transparency and interactivity in recommender systems, continuously refining performance through user feedback. Despite challenges, researchers propose approaches to effectively leverage LLMs, aiming to bridge the gap between strengths and challenges for improved system performance.
Essentially, this paper makes three significant contributions to the realm of LLMs in recommenders:
\begin{itemize}
    \item Introducing a systematic taxonomy designed to categorize LLMs for recommenders.
    \item Systematizing the essential and primary techniques illustrating how LLMs are utilized in recommender systems, providing a detailed overview of current research in this domain.
    \item Deliberating on the challenges and limitations associated with traditional recommender systems, accompanied by solutions using LLMs in recommenders.
\end{itemize}

\begin{figure*}[h!]
  \centering
  \resizebox{0.82\textwidth}{!}{
    \begin{forest}
    forked edges,
      for tree={
        grow=east,
        reversed=true,
        anchor=base west,
        parent anchor=east,
        child anchor=west,
        base=center,
        font=\large,
        rectangle,
        draw=hidden-draw,
        rounded corners,
        align=center,
        text centered,
        minimum width=5em,
        edge+={darkgray, line width=1pt},
        s sep=3pt,
        inner xsep=2pt,
        inner ysep=3pt,
        line width=0.8pt,
        ver/.style={rotate=90, child anchor=north, parent anchor=south, anchor=center},
      },
      where level=1{text width=15em,font=\normalsize,}{},
      where level=2{text width=14em,font=\normalsize,}{},
      where level=3{minimum width=10em,font=\normalsize,}{},
      where level=4{text width=26em,font=\normalsize,}{},
      where level=5{text width=20em,font=\normalsize,}{},
      [
        \textbf{LLMs in Recommender Systems}\\ \S\ref{sec:LLMs reccommendation}, for tree={fill=paired-light-red!70},text width = 18em
                [
        \textbf{LLM-Powered Recommender Systems}\\ \S\ref{subsec:content receommender}, fill=bg4,text width = 22em
        [\textbf{LlamaRec} \cite{yue2023llamarec}\\
        \textbf{CoLLM} \cite{zhang2023collm}\\
        \textbf{RecMind} \cite{wang2023llm4vis}\\
        \textbf{RecRec} \cite{Verma_2023}\\
        \textbf{P5} \cite{geng2023recommendation}\\
        \textbf{RecExplainer} \cite{lei2023recexplainer}\\
        \textbf{DOKE} \cite{yao2023knowledge}\\
        \textbf{RLMRec} \cite{ren2024representation}\\
        \textbf{RARS} \cite{Di}\\
        \textbf{GenRec} \cite{ji2023genrec}\\
        \textbf{RIF} \cite{zhang2024tsrankllm}\\
        \textbf{Recommender AI Agent} \cite{huang}\\
        \textbf{POSO} \cite{du2023enhancing}, fill=bg29, text width = 22em]
        ]
        [\textbf{Off-the-shelf  Recommender Systems}\\ \S\ref{subsec:notunning}, fill=bg15,text width = 22em
        [\textbf{RecAgent} \cite{wang2023large}\\
        \textbf{AnyPredict} \cite{wang2023meditab}\\
        \textbf{ZRRS} \cite{Hou}\\
        \textbf{LGIR} \cite{du2023enhancing}\\
        \textbf{MINT} \cite{mysore2023large}\\
        \textbf{LLM4Vis} \cite{wang2023llm4vis}\\
        \textbf{ONCE} \cite{liu2023once}\\
        \textbf{GPT4SM} \cite{peng}\\
        \textbf{TransRec} \cite{lin2023multifacet}\\
        \textbf{Agent4Rec} \cite{Zhang_2023}\\
        \textbf{Collaborative LLMs} \cite{zhu2023collaborative}, fill=bg40,text width = 22em
        ]
        ]
        [\textbf{Sequential Recommender Systems}\\ \S\ref{subsec:Sequential Recommendations}, fill=bg33, text width = 22em
        [\textbf{PDRec} \cite{ma2024plugin}\\
        \textbf{G-Meta} \cite{xiao2024gmeta}\\
        \textbf{ELCRec} \cite{liu2024endtoend}\\
        \textbf{GPTRec} \cite{petrov2023generative}\\
        \textbf{DRDT} \cite{wang2023drdt}\\
        \textbf{LLaRA} \cite{liao2023llara}\\
        \textbf{E4SRec} \cite{li2023e4srec}\\
        \textbf{RecInterpreter} \cite{yang2023large}\\
        \textbf{VQ-Rec} \cite{hou2024large}\\
        \textbf{One Model for All} \cite{tang2023model},fill=bg22,text width = 22em 
        ]
        ]
        [\textbf{Conversational Recommender Systems}\\ \S\ref{subsec:LLMs in Conversational Recommender}, fill=bg10, text width = 22em
        [\textbf{CRSs} \cite{sun2018conversational}\\
        \textbf{LLMCRS} \cite{feng2023large}\\
        \textbf{Chat-REC} \cite{gao2023chatrec}\\
        \textbf{ChatQA} \cite{liu2024chatqa},fill=bg8,text width = 22em 
        ]
        ]
        [\textbf{Personalized Recommender Systems}\\ \S\ref{subsec:Personalized Recommender}, fill=bg21, text width = 22em
        [\textbf{PALR} \cite{yang2023palr}\\
        \textbf{Enhanced Recommendation} \cite{zhang2023recommendation}\\
        \textbf{PAP-REC} \cite{li2024paprec}\\
        \textbf{Health-LLM} \cite{jin2024healthllm}\\
        \textbf{Music Recommender} \cite{briand2024lets}\\
        \textbf{ControlRec} \cite{qiu2023controlrec}, fill=bg5,text width = 22em 
        ]
        ]
        [\textbf{Knowledge Graph-enhanced Recommender Systems}\\ \S\ref{subsec:Knowledge Graph}, fill=bg1, text width = 22em
        [\textbf{KoLA} \cite{wang2024enhancing}\\
        \textbf{KAR} \cite{xi2023openworld}\\
        \textbf{LLMRG} \cite{Wang_2019},fill=bg121,text width = 22em 
        ]
        ]
        [
        \textbf{Reranking in Recommender Systems}\\ \S\ref{subsec:Diverse rerenaking}, for tree={fill=bg39}, text width = 22em
        [
        \textbf{Diverse Reranking} \cite{carraro2024enhancing}\\
        \textbf{Multishot Reranker} \cite{xiao2024multislot}\\
        \textbf{Ranking GPT} \cite{zhang2024tsrankllm}, fill = bg25,text width = 22em 
        ]
        ]
        [
        \textbf{Prompt Engineered LLMs in Recommender Systems}\\ \S\ref{subsec: promptenginnering}, for tree={fill=bg5}, text width = 22em
        [
        \textbf{Reprompting} \cite{spurlock2024chatgpt}\\
        \textbf{ProLLM4Rec} \cite{xu2024prompting}\\
        \textbf{UEM} \cite{doddapaneni2024user}\\
        \textbf{POD} \cite{li2024paprec}\\
        \textbf{M6-Rec} \cite{cui2022m6rec}\\
        \textbf{PBNR} \cite{li2024paprec}\\
        \textbf{LLMRG} \cite{wang}\\
        \textbf{RecSys} \cite{fan2023recommender}, fill = bg113,text width = 22em 
        ]
        ]
        [
        \textbf{Fine-tuned LLMs for Recommender Systems}\\ \S\ref{subsec:finetuned LLMs}, for tree={fill=bg40}, text width = 22em
        [
        \textbf{TALLRec} \cite{Bao}\\
        \textbf{Flan-T5} \cite{kang2023llms}\\
        \textbf{InstrucRec} \cite{zhang2023bridging}\\
        \textbf{RecLLM} \cite{friedman2023leveraging}\\
        \textbf{DEALRec} \cite{lin2024dataefficient}\\
        \textbf{INTERS} \cite{zhu2024inters}, fill = bg10,text width = 22em 
        ]
        ]
        [
        \textbf{Evaluating LLMs in Recommender Systems}\\ \S\ref{section:evaluation}, for tree={fill=bg115}, text width = 22em
        [
        \textbf{iEvaLM} \cite{wang}\\
        \textbf{FaiRLLM} \cite{zhang2024tsrankllm}\\
        \textbf{Ranking GPT} \cite{Zhang_2023}, fill=pg69,text width = 22em 
        ]
        ]
      ]
    \end{forest}
  }
  \caption{Taxonomy of Recommendation in LLMs, encompassing LLMs in Recommendation Systems, Sequential \& Conversational Recommender Systems, Personalized recommender system Knowledge Graph enhancements, Reranking, Prompts, and Fine-Tuned LLMs. Offers a condensed overview of models and methods within the Recommendation landscape.}
  \label{fig:lit_surv}
\end{figure*}
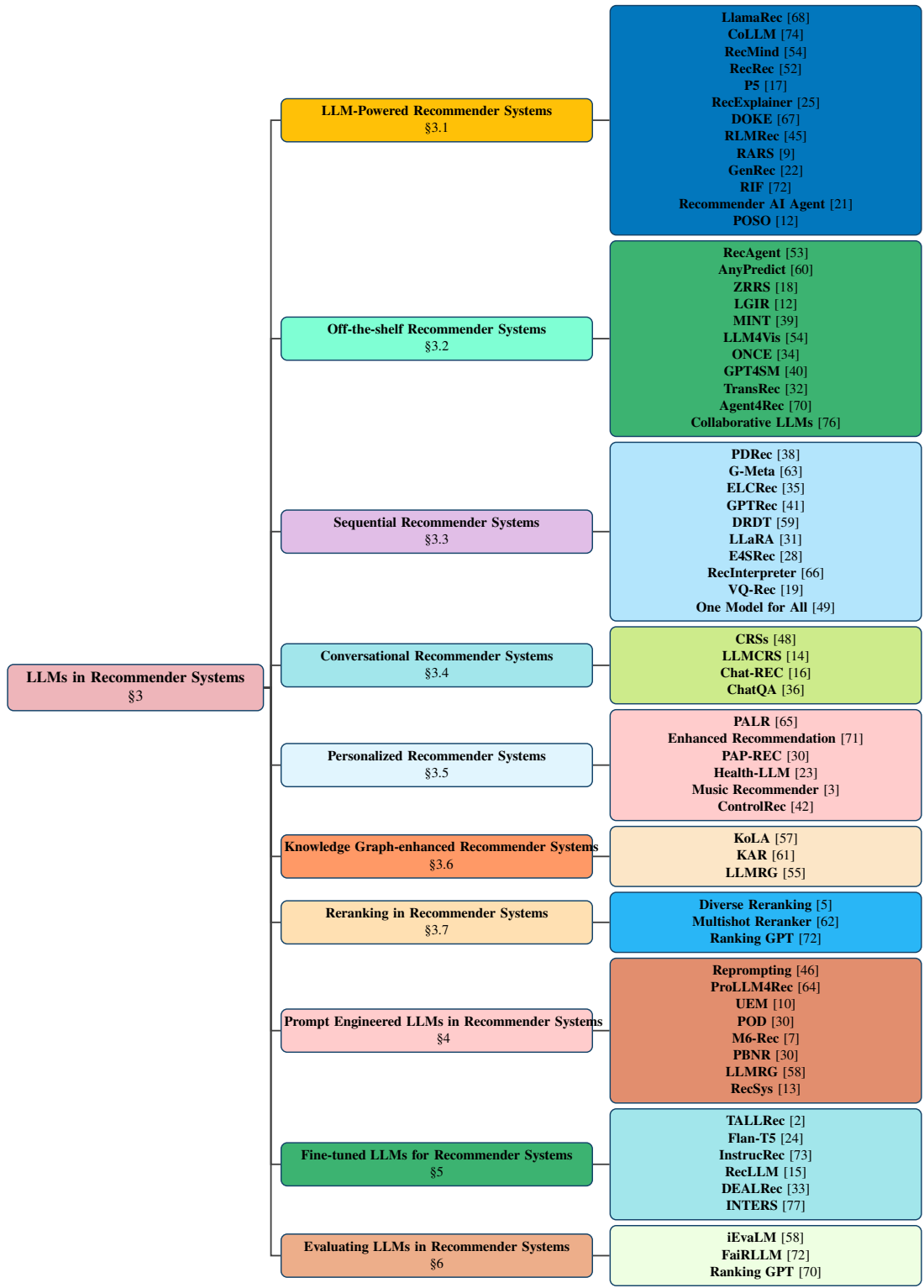

\section{Background and Related Work}

\label{sec:Backgroud}
In this section, we provide a concise overview of pertinent literature concerning recommender systems and methods involving LLMs.
\subsection{Recommender Systems}
Traditional recommender systems follow Candidate Generation, Retrieval, and Ranking phases. However, the advent of LLMs brings a new perspective. Unlike conventional models, LLMs do not require separate embeddings for each user/item interaction. Instead, they use task-specific prompts encompassing user data, item information, and previous preferences. This adaptability allows LLMs to generate recommendations directly, dynamically adapting to various contexts without explicit embeddings. While departing from traditional models, this unified approach retains the capacity for personalized and contextually-aware recommendations, offering a more cohesive and adaptable alternative to segmented retrieval and ranking structures.

\subsection{Large Language Models (LLMs)}
LLMs, such as Llama \cite{touvron2023llama}, GPT ~\cite{radford2018improving}, T5~\cite{raffel}, etc., are versatile in NLP. BERT is an encoder-only model with bidirectional attention, GPT employs a transformer decoder for one-directional processing, and T5 transforms NLP problems into text generation tasks. Recent LLMs like GPT-3~\cite{brown},  Language Model for Dialogue Applications (LaMDA)~\cite{thoppilan}, Pathways Language Model (PaLM)~\cite{chowdhery}, and Vicuna excel in understanding human-like textual knowledge, employing In-Context Learning (ICL)~\cite{dong2023survey} for context-based responses. 

\section{LLMs in Recommender Systems}
\label{sec:LLMs reccommendation}
In this section, we will investigate how LLMs enhance deep learning-based recommender systems by playing crucial roles in user data collection, feature engineering, and scoring/ranking functions. Going beyond their role as mere components, LLMs actively govern the system pipeline, fostering interactive and explainable recommendation processes. They possess the capability to comprehend user preferences, orchestrate ranking stages, and contribute to the evolution of integrated conversational recommender systems.\\
\subsection{LLM-Powered Recommender Systems Across Domains}
\label{subsec:content receommender}
In this section, we delve into the wide-reaching applications of LLM-powered recommender systems across diverse domains. From entertainment to shopping and conversational agents, witness the versatile impact of LLMs.\\
\textbf{LlamaRec}: Yue~\etal~\cite{yue2023llamarec} introduce LlamaRec, a two-stage recommendation system. In the first stage, a sequential recommender uses user history to efficiently choose candidate items. The selected candidates and user history are then fed into an LLM using a tailored prompt template. Instead of autoregressive generation, a verbalizer is used to convert LLM output logits into probability distributions, speeding up the inference process and enhancing efficiency. This novel approach overcomes the text generation sluggishness typically encountered, resulting in more efficient recommendations.\\
\textbf{RecMind}: Wang ~\etal~\cite{wang} introduce RecMind, an innovative recommender agent fueled by LLMs. RecMind, as an autonomous agent, is intricately designed to furnish personalized recommendations through strategic planning and the utilization of external tools. RecMind incorporates the Self-Inspiring (SI) planning algorithm. This algorithm empowers the agent by enabling it to consider all previously explored planning paths, thereby enhancing the generation of more effective recommendations. The agent's comprehensive framework includes planning, memory, and tools, collectively bolstering its capacities in reasoning, acting, and memory retention.\\
\textbf{RecRec}: Verma~\etal~\cite{Verma_2023} introduce RecRec, aiming to create algorithmic recourse-based explanations for content filtering-based recommender systems. RecRec offers actionable insights, suggesting specific actions to modify recommendations based on desired preferences. The authors advocate for an optimization-based approach, validating RecRec's effectiveness through empirical evaluations on real-world datasets. The results demonstrate its ability to generate valid, sparse, and actionable recourses that provide valuable insights for improving product rankings.\\
\textbf{P5}: Geng~\etal~\cite{geng2023recommendation} introduce a groundbreaking contribution made by introducing a unified ``Pretrain, Personalized Prompt \& Predict Paradigm." This paradigm seamlessly integrates various recommendation tasks into a cohesive conditional language generation framework. It involves the creation of a carefully designed set of personalized prompts covering five distinct recommendation task families.Noteworthy is P5's robust zero-shot generalization capability, demonstrating its effectiveness in handling new personalized prompts and previously unseen items in unexplored domains.\\
\textbf{RecExplainer}: Lei~\etal~\cite{lei2023recexplainer} introduce utilization of LLMs as alternative models for interpreting and elucidating recommendations from embedding-based models, which often lack transparency. The authors introduce three methods for aligning LLMs with recommender models: behavior alignment, replicating recommendations using language; intention alignment, directly comprehending recommender embeddings; and hybrid alignment, combining both approaches. LLMs can adeptly comprehend and generate high-quality explanations for recommendations, thereby addressing the conventional tradeoff between interpretability and complexity.\\
\textbf{DOKE}: Yao~\etal~\cite{yao2023knowledge}, introduce DOKE, a paradigm enhancing LLMs with domain-specific knowledge for practical applications. DOKE utilizes an external domain knowledge extractor to prepare and express knowledge in LLM-understandable formats, avoiding costly fine-tuning. Demonstrated in recommender systems, DOKE provides item attributes and collaborative filtering signals.\\
\textbf{RLMRec}: Ren~\etal~\cite{ren2024representation} introduce recommender system challenges related to graph-based models and ID-based data limitations. Introducing RLMRec, a model-agnostic framework, the paper integrates LLMs to improve recommendations by leveraging representation learning and aligning semantic spaces. The study establishes a theoretical foundation and practical evaluations demonstrate RLMRec's robustness to noise and incomplete data in enhancing the recommendation process.\\
\textbf{RARS}: Di~\etal~\cite{Di} highlights that LLMs exhibit contextual awareness and a robust ability to adapt to previously unseen data.By amalgamating these technologies, a potent tool is formed for delivering contextual and pertinent recommendations, particularly in cold scenarios marked by extensive data sparsity. It introduces a innovative approach named Retrieval-augmented Recommender Systems, which merges the advantages of retrieval-based and generation-based models to augment the capability of RSs in offering pertinent suggestions.\\
\textbf{GenRec}: Ji~\etal~\cite{ji2023genrec} intorduces a new application of LLMs in recommendation systems, particularly enhancing user engagement with diverse data. The paper introduces GenRec model which leverages descriptive item information, leading to more sophisticated personalized recommendations. Experimental results confirms GenRec's effectiveness, indicating its potential for various applications and encouraging further research on LLMs in generative recommender systems. \\
\textbf{Recommendation as Instruction Following}: Zhang~\etal~\cite{zhang2023recommendation} introduce a novel recommendation concept by expressing user preferences through natural language instructions for LLMs. It involves fine-tuning a 3B Flan-T5-XL LLM to align with recommender systems, using a comprehensive instruction format. This paradigm treats recommendation as instruction-following, allowing users flexibility in expressing diverse information needs.\\
\textbf{Recommender AI Agent}: Huang~\etal~\cite{huang} introduce RecAgent, a pioneering framework merging LLMs and recommender models for an interactive conversational recommender system. Addressing the strengths and weaknesses of each, RecAgent uses LLMs for language comprehension and reasoning (the 'brain') and recommender models for item recommendations (the 'tools'). The paper outlines essential tools, including a memory bus for communication, dynamic demonstration-augmented task planning, and a reflection strategy for quality evaluation, creating a flexible and comprehensive system.\\
\textbf{CoLLM}: Zhang~\etal~\cite{zhang2023collm} introduce LLMRec, emphasizing collaborative information modeling alongside text semantics in recommendation systems. CoLLM, seamlessly incorporating collaborative details into LLMs using external traditional models for improved recommendations in cold and warm start scenarios.\\
\textbf{POSO}: Dai~\etal~\cite{dai2021poso} introduce a Personalized COld Start MOdules (POSO),enhancing pre-existing modules with user-group-specialized sub-modules and personalized gates for comprehensive representations. Adaptable to various modules like Multi-layer Perceptron and Multi-head Attention, POSO shows significant performance improvement with minimal computational overhead.

\subsection{Off-the-shelf LLM-based Recommender Systems}
\label{subsec:notunning}
In this section we examine recommender systems that operate without tuning, specifically focusing on whether adjustments have been made to the LLM.\\
\textbf{RecAgent}: Wang~\etal~\cite{wang2023large} proposes the potential of models for robust user simulation, particularly in reshaping traditional user behavior analysis. They focus on recommender systems, employing LLMs to conceptualize each user as an autonomous agent within a virtual simulator named RecAgent. It introduces global functions for real-human playing and system intervention, enhancing simulator flexibility. Extensive experiments are conducted to assess the simulator's effectiveness from both agent and system perspectives.\\
\textbf{MediTab}: Wang~\etal~\cite{wang2023meditab} introduce MediTab, a method enhancing scalability for medical tabular data predictors across diverse inputs. Using an LLM-based data engine, they merge tabular samples with distinct schemas. Through a "learn, annotate, and refinement" pipeline, MediTab aligns out-domain data with the target task, enabling it to make inferences for arbitrary tabular inputs without fine-tuning. Achieving impressive results on patient and trial outcome prediction datasets, MediTab demonstrates substantial improvements over supervised baselines and outperforms XGBoost models in zero-shot scenarios.\\
\textbf{ZRRS}: Hou~\etal~\cite{hou2024large} introduce a recommendation problem as a conditional ranking task, using LLMs to address it with a designed template. Experimental results on widely-used datasets showcase promising zero-shot ranking capabilities.The authors propose special prompting and bootstrapping strategies, demonstrating effectiveness. These insights position zero-shot LLMs as competitive challengers to conventional recommendation models, particularly in ranking candidates from multiple generators.\\
~\textbf{LGIR}: Du~\etal~\cite{du2023enhancing} proposes an innovative job recommendation method based on LLMs that overcomes limitations in fabricated generation. This comprehensive approach enhances the accuracy and meaningfulness of resume completion, even for users with limited interaction records. To address few-shot problems, the authors suggest aligning low-quality resumes with high-quality generated ones using Generative Adversarial Networks (GANs), refining representations and improving recommendation outcomes.\\
\textbf{MINT}: Mysore~\etal~\cite{mysore2023large} use LLMs for data augmentation in training Narrative-Driven Recommendation (NDR) models. LLMs generate synthetic narrative queries from user-item interactions using few-shot prompting. Retrieval models for NDR are trained with a combination of synthetic queries and original interaction data. Experiments show the approach's effectiveness, making it a successful strategy for training compact retrieval models that outperform alternatives and LLM baselines in narrative-driven recommendation.\\
\textbf{PEPLER}: Li ~\etal~\cite{li2023personalized} introduce PEPLER, a system that generates natural language explanations for recommendations by treating user and item IDs as prompts. Two training strategies are proposed to bridge the gap between continuous prompts and the pre-trained model, aiming to enhance the performance of explanation generation. The researchers suggest that this approach could inspire others in tuning pre-trained language models more effectively. Evaluation of the generated explanations involves not only text quality metrics such as BLEU and ROUGE but also metrics focused on explainability from the perspective of item features. Results from extensive experiments indicate that PEPLER consistently outperforms state-of-the-art baselines.\\
\textbf{LLM4Vis}: Wang~\etal~\cite{wang2023llm4vis} introduce LLM4Vis, a ChatGPT-based method for accurate visualization recommendations and human-like explanations with minimal demonstration examples. The methodology involves feature description, demonstration example selection, explanation generation, demonstration example construction, and inference steps. A new explanation generation bootstrapping method refines explanations iteratively, considering previous generations and using template-based hints.\\
\textbf{ONCE}: Liu~\etal~\cite{liu2023once} introduce a combination of open-source and closed-source LLMs to enhance content-based recommendation systems . Open-source LLMs contribute as content encoders, while closed-source LLMs enrich training data using prompting techniques. Extensive experiments show significant effectiveness, with a relative improvement of up to 19.32\% compared to existing models, highlighting the potential of both LLM types in advancing content-based recommendations.\\
\textbf{GPT4SM}: Peng~\etal~\cite{peng}proposes three strategies to integrate the knowledge of LLMs into basic PLMs, aiming to improve their overall performance. These strategies involve utilizing GPT embedding as a feature (EaaF) to enhance text semantics, using it as a regularization (EaaR) to guide the aggregation of text token embeddings, and incorporating it as a pre-training task (EaaP) to replicate the capabilities of LLMs. The experiments conducted by the researchers demonstrate that the integration of GPT embeddings enables basic PLMs to enhance their performance in both advertising and recommendation tasks.\\
\textbf{TransRec}: Lin~\etal~\cite{lin2023multifacet} introduce TransRec, a pioneering multi-facet paradigm designed to establish a connection between LLMs and recommendation systems. TransRec employs multi-facet identifiers, including ID, title, and attribute information, to achieve a balance of distinctiveness and semantic richness. Additionally, the authors present a specialized data structure for TransRec, ensuring precise in-corpus identifier generation. The adoption of substring indexing is implemented to encourage LLMs to generate content from various positions. The researchers conduct the implementation of TransRec on two core LLMs, specifically BART-large and LLaMA-7B.\\
\textbf{Agent4Rec}: Zhang~\etal~\cite{zhang2023generative} introduce Agent4Rec, a movie recommendation simulator using LLM-empowered generative agents. These agents have modules for user profiles, memory, and actions tailored for recommender systems. The study explores how well LLM-empowered generative agents simulate real human behavior in recommender systems, evaluating alignment and deviations between agents and user preferences. Experiments delve into the filter bubble effect and uncover causal relationships in recommendation tasks.\\
\textbf{Collaborative LLMs for Recommender Systems}: Zhu~\etal~\cite{zhu2023collaborative} introduce a Collaborative LLMscite, as a novel recommendation system. It combines pretrained LLMs with traditional ones to address challenges in spurious correlations and language modeling. The methodology extends the LLM's vocabulary with special tokens for users and items, ensuring accurate user-item interaction modeling. Mutual regularization in pretraining connects collaborative and content semantics, and stochastic item reordering manages non-crucial item order.
\subsection{LLMs in Sequential Recommender Systems}
\label{subsec:Sequential Recommendations}
It is a recommendation approach that focuses on the order of a user's past interactions to predict their next likely actions. It considers the sequence of items a user has viewed, purchased, or interacted with, rather than just overall popularity or static preferences. This allows for more personalized and timely recommendations that reflect the user's current interests and evolving tastes.\\
\textbf{PDRec}: Ma~\etal~\cite{ma2024plugin} proposes a Plug-in Diffusion Model for Sequential Recommendation. This innovative framework employs diffusion models as adaptable plugins, capitalizing on user preferences related to all items generated through the diffusion process to address the challenge of data sparsity. By incorporating time-interval diffusion, PDRec infers dynamic user preferences, adjusts historical behavior weights, and enhances potential positive instances. Additionally, it samples noise-free negatives from the diffusion output to optimize the model.\\
\textbf{G-Meta}: Xio~\etal~\cite{xiao2024gmeta} introduce G-Meta framework, which is designed for the distributed training of optimization-based meta-learning recommendation models on GPU clusters. It optimizes computation and communication efficiency through a combination of data and model parallelism, and includes a Meta-IO pipeline for efficient data ingestion. Experimental results demonstrate G-Meta's ability to achieve accelerated training without compromising statistical performance. Since 2022, it has been implemented in Alibaba's core systems, resulting in a notable 4x reduction in model delivery time and significant improvements in business metrics. Its key contributions include hybrid parallelism, distributed meta-learning optimizations, and the establishment of a high-throughput Meta-IO pipeline.\\
\textbf{ELCRec}: Liu~\etal~\cite{liu2024endtoend} presents ELCRec, a novel approach for intent learning in sequential recommendation systems. ELCRec integrates representation learning and clustering optimization within an end-to-end framework to capture users' intents effectively. It uses learnable parameters for cluster centers, incorporating clustering loss for concurrent optimization on mini-batches, ensuring scalability. The learned cluster centers serve as self-supervision signals, enhancing representation learning and overall recommendation performance.\\
\textbf{GPTRec}: Petrov~\etal~\cite{petrov2023generative} introduce GPTRe, a GPT-2-based sequential recommendation model using the SVD Tokenisation algorithm to address vocabulary challenges. The paper presents a Next-K recommendation strategy, demonstrating its effectiveness. Experimental results on the MovieLens-1M dataset show that GPTRec matches SASRec's quality while reducing the embedding table by 40\%.\\
\textbf{DRDT}: Wang~\etal~\cite{wang2023drdt} introduce the improvement of LLM reasoning in sequential recommendations. The methodology introduces Dynamic Reflection with Divergent Thinking (DRDT), within a retriever-reranker framework. Leveraging a collaborative demonstration retriever, it employs divergent thinking to comprehensively analyze user preferences. The dynamic reflection component emulates human learning, unveiling the evolution of users' interests.\\
\textbf{LLaRA}: Liao~\etal~\cite{liao2023llara} introduce LLaRA, a framework for modeling sequential recommendations within LLMs. LLaRA adopts a hybrid approach, combining ID-based item embeddings from conventional recommenders with textual item features in LLM prompts. Addressing the "sequential behavior of the user" as a novel modality, an adapter bridges the gap between traditional recommender ID embeddings and LLM input space. Utilizing curriculum learning, researchers gradually transition from text-only to hybrid prompting during training, enabling LLMs to adeptly handle sequential recommendation tasks.\\
\textbf{E4SRec}: Li~\etal~\cite{li2023e4srec} introduce E4SRec, seamlessly integrating LMs with traditional recommender systems using item IDs. E4SRec efficiently generates ranking lists in a single forward process, addressing challenges in integrating IDs with LLMs and proposing an industrial-level recommender system with demonstrated superiority in real-world applications.\\
\textbf{RecInterpreter}: Yang~\etal~\cite{yang2023large} propose RecInterpreter, evaluating LLMs for interpreting sequential recommender representation space. Using multimodal pairs and lightweight adapters, RecInterpreter enhances LLaMA's understanding of ID-based sequential recommenders, particularly with sequence-residual prompts. It also enables LLaMA to determine the ideal next item for a user from generative recommenders like DreamRec\\
\textbf{VQ-Rec}: Hou~\etal~\cite{Hou} propose VQ-Rec, a novel method for transferable sequential recommenders using Vector-Quantized item representations. It translates item text into indices, generates representations, and employs enhanced contrastive pre-training with mixed-domain code representations. A differentiable permutation-based network guides a unique cross-domain fine-tuning approach, demonstrating effectiveness across six benchmarks in various scenarios.\\
\textbf{K-LaMP}: Baek~\etal~\cite{baek2023knowledgeaugmented} proposes enhancing LLMs by incorporating user interaction history with a search engine for personalized outputs. The study introduces an entity-centric knowledge store derived from users' web search and browsing activities, forming the basis for contextually relevant LLM prompt augmentations.\\
\textbf{One Model for All}: Tang~\etal~\cite{tang2023model} presents LLM-Rec, addressing challenges in multi-domain sequential recommendation. They use an LLM to capture world knowledge from textual data, bridging gaps between recommendation scenarios. The task is framed as a next-sentence prediction task for the LLM, representing items and users with titles. Increasing the pre-trained language model size enhances both fine-tuned and zero-shot domain recommendation, with slight impacts on fine-tuning performance. The study explores different fine-tuning methods, noting performance variations based on model size and computational resources.
\subsection{LLMs in Conversational Recommender Systems}
\label{subsec:LLMs in Conversational Recommender}
This section explores the role of LLMs in Conversational Recommender Systems (CRSs)~\cite{sun2018conversational}. CRSs aim to provide quality recommendations through dialogue interfaces, covering tasks like user preference elicitation, recommendation, explanation, and item information search. However, challenges arise in managing sub-tasks, ensuring efficient problem-solving, and generating user-interaction-friendly responses in the development of effective CRS.\\
\textbf{LLMCRS}: Feng~\etal~\cite{feng2023large} introduce LLMCRS, a pioneering LLM-based Conversational Recommender System. It strategically uses LLM for sub-task management, collaborates with expert models, and leverages LLM's generation capabilities. LLMCRS incorporates instructional mechanisms and proposes fine-tuning with reinforcement learning from CRSs performance feedback (RLPF) for conversational recommendations. Experimental findings show LLMCRS with RLPF outperforms existing methods, demonstrating proficiency in handling conversational recommendation tasks.\\
\textbf{Chat-REC}: Gao~\etal~\cite{gao2023chatrec} introduce Chat-REC which uses an LLM to enhance its conversational recommender by summarizing user preferences from profiles. The system combines traditional recommendation methods with OpenAI's Chat-GPT for multi-round recommendations, interactivity, and explainability. To handle cold item recommendations, an item-to-item similarity approach using external current information is proposed. In experiments, Chat-REC performs well in zero-shot and cross-domain recommendation tasks.\\
\textbf{ChatQA}: Liu~\etal~\cite{liu2024chatqa} introduce ChatQA conversational question-answering models rivaling GPT-4's performance, without synthetic data. They propose a two-stage instruction tuning method to enhance zero-shot capabilities. Demonstrating the effectiveness of fine-tuning a dense retriever on multi-turn QA datasets, they show it matches complex query rewriting models with simpler deployment.
\subsection{LLMs in Personalized Recommenders Systems}
\label{subsec:Personalized Recommender}
\textbf{PLAR}: Yang~\etal~\cite{yang2023palr} introduce PALR, a versatile personalized recommendation framework addressing challenges in the domain. The process involves utilizing an LLM and user behavior to generate user profile keywords, followed by a retrieval module for candidate pre-filtering. PALR, independent of specific retrieval algorithms, leverages the LLM to provide recommendations based on user historical behaviors. To tailor general-purpose LLMs for recommendation scenarios, user behavior data is converted into prompts, and an LLaMa 7B model is fine-tuned. PALR demonstrates competitive performance against state-of-the-art methods on two public datasets, making it a flexible recommendation solution. \\
\textbf{Bridging LLMs and Domain-Specific Models for Enhanced Recommendation}: Zhang~\etal~\cite{zhang2023bridging} introduce information disparity between domain-specific models and LLMs for personalized recommendation. The approach incorporates an information sharing module, acting as a repository and conduit for collaborative training. This enables a reciprocal exchange of insights: domain models provide user behavior patterns, and LLMs contribute general knowledge and reasoning abilities. Deep mutual learning during joint training enhances this collaboration, bridging information gaps and leveraging the strengths of both models. \\
\textbf{PAP-REC}: Li~\etal~\cite{li2024paprec} proposes PAP-REC a framework that automatically generates personalized prompts for recommendation language models (RLMs) to enhance their performance in diverse recommendation tasks. Instead of depending on inefficient and suboptimal manually designed prompts, PAP-REC utilizes gradient-based methods to effectively explore the extensive space of potential prompt tokens. This enables the identification of personalized prompts tailored to each user, contributing to improved RLM performance. \\
\textbf{Health-LLM}: Jin~\etal~\cite{jin2024healthllm} introduce a framework integrating LLM and medical expertise for enhanced disease prediction and health management using patient health reports. Health-LLM involves extracting informative features, assigning weighted scores by medical professionals, and training a classifier for personalized predictions. Health-LLM offers detailed modeling, individual risk assessments, and semi-automated feature engineering, providing professional and tailored intelligent healthcare.\\
\textbf{Personalized Music Recommendation}: Brain~\etal~\cite{briand2024lets} introduce a Deezer's shift to a fully personalized system for improved new music release discoverability. The use of cold start embeddings and contextual bandits leads to a substantial boost in clicks and exposure for new releases through personalized recommendations.\\
\textbf{GIRL}: Zheng~\etal~\cite{zheng2023generative} introduce GIRL, a job recommendation approach inspired by LLMs. It uses Supervised Fine-Tuning (SFT) for generating Job Descriptions (JDs) from CVs, incorporating a reward model for CV-JD matching. Reinforcement Learning fine-tunes the generator with Proximal Policy Optimization, creating a candidate-set-free, job seeker-centric model. Experiments on a real-world dataset demonstrate significant effectiveness, signaling a paradigm shift in personalized job recommendation.\\
\textbf{ControlRec}: Qiu~\etal~\cite{qiu2023controlrec} introduce ControlRec, a framework for contrastive prompt learning integrating LLMs into recommendation systems. User/item IDs and natural language prompts are treated as heterogeneous features and encoded independently. The framework introduces two contrastive objectives: Heterogeneous Feature Matching (HFM), aligning item descriptions with IDs based on user interactions, and Instruction Contrastive Learning (ICL), merging ID and language representations by contrasting output distributions for recommendation tasks.
\subsection{LLMs for Knowledge Graph-enhanced Recommender Systems}
\label{subsec:Knowledge Graph}
This section explores how Language Models (LLMs) are utilized to enhance knowledge graphs within recommender systems, leveraging natural language understanding to enrich data representation and recommendation outcomes.\\
\textbf{KoLA}: Wang~\etal~\cite{wang2024enhancing} introduce the significance of incorporating knowledge graphs in recommender systems, as indicated by the benchmark evaluation KoLA~\cite{yue2023llamarec}. Knowledge graphs enhance recommendation explainability, with embedding-based, path-based, and unified methods improving recommendation quality. Challenges in dataset sparsity, especially in private domains, are noted, and LLMs like symbolic-kg help address data scarcity, though issues like the phantom problem and common sense limitations must be tackled for optimal recommender system performance\\
\textbf{KAR}: Xi~\etal~\cite{xi2023openworld} introduce KAR, the Open-World Knowledge Augmented Recommendation Framework with LLM. KAR extracts reasoning knowledge on user preferences and factual knowledge on items from LLMs using factorization prompting. The resulting knowledge is transformed into augmented vectors using a hybrid-expert adaptor, enhancing the performance of any recommendation model. Efficient inference is ensured by preprocessing and prestoring LLM knowledge.\\
\textbf{LLMRG}: LLM Reasoning Graphs (LLMRG) by Wang~\etal~\cite{wang2024enhancing} utilize LLMs to create personalized reasoning graphs for robust and interpretable recommender systems. LLMRG connects user profiles and behavioral sequences, incorporating chained graph reasoning, divergent extension, self-verification, and self-improvement of knowledge bases. LLMRG enhances recommendations without extra user or item information, offering adaptive reasoning for a comprehensive understanding of user preferences and behaviors.

\subsection{LLMs for Reranking in Recommender Systems}
\label{subsec:Diverse rerenaking}
This section focuses on the utilization of LLMs for reranking within recommendation systems, emphasizing their role in enhancing the overall recommendation process.\\
\textbf{Diverse Reranking}: Carraro~\etal~\cite{carraro2024enhancing} introduce LLMs for reranking recommendations to enhance diversity beyond mere relevance. In an initial study, the authors confirm LLMs' ability to proficiently interpret and perform reranking tasks with a focus on diversity. They then introduce a more rigorous methodology, prompting LLMs in a zero-shot manner to create a diverse reranking from an initial candidate ranking generated by a matrix factorization recommender.\\
\textbf{MultiSlot ReRanker}: Xiao~\etal~\cite{xiao2024multislot} proposes MultiSlot, ReRanker, a model-based framework for re-ranking in recommendation systems, addressing relevance, diversity, and freshness. It employs the efficient Sequential Greedy Algorithm and utilizes an OpenAI Gym simulator to evaluate learning algorithms for re-ranking under diverse assumptions.\\
\textbf{Zero-Shot Ranker}: Hou~\etal~\cite{hou2024large} introduce a recommendation as a conditional ranking task, utilizing LLMs with carefully designed prompts based on sequential interaction histories. Despite promising zero-shot ranking abilities, challenges include accurately perceiving interaction order and susceptibility to popularity biases. The study suggests that specially crafted prompting and bootstrapping strategies can alleviate these issues, enabling zero-shot LLMs to competitively rank candidates from multiple generators compared to traditional recommendation models.\\
\textbf{RankingGPT}: Zhang~\etal~\cite{zhang2024tsrankllm} introduce a two-stage training strategy to enhance LLMs for efficient text ranking. The initial stage involves pretraining the LLM on a vast weakly supervised dataset, aiming to improve its ability to predict associated tokens without altering its original objective. Subsequently, supervised fine-tuning is applied with constraints to enhance the model's capability in discerning relevant text pairs for ranking while preserving its text generation abilities.

\section{Prompt Engineered LLMs in Recommender Systems}
\label{subsec: promptenginnering}
This section investigates the incorporation of prompt engineering by LLMs in the context of recommendation systems, emphasizing their role in refining and enhancing user prompts for improved recommendations.\\
\textbf{Reprompting}: Spur~\etal~\cite{spurlock2024chatgpt} introduce ChatGPT as a conversational recommendation system, emphasizing realistic user interactions and iterative feedback for refining suggestions. The study investigates popularity bias in ChatGPT's recommendations, highlighting that iterative reprompting significantly improves relevance. Results show ChatGPT outperforms random and traditional systems, suggesting effective mitigation of popularity bias through strategic prompt engineering.\\
\textbf{ProLLM4Rec}: Xu~\etal~\cite{xu2024prompting} introduce ProLLM4Rec, as a comprehensive framework utilizing LLMs as recommender systems through prompting engineering. The framework focuses on selecting the LLM based on factors like availability, architecture, scale, and tuning strategies, and crafting effective prompts that include task descriptions, user modeling, item candidates, and prompting strategies.\\
\textbf{UEM}: Doddapaneni~\etal~\cite{doddapaneni2024user} introduce User Embedding Module (UEM), This module is designed to effectively handle extensive user preference histories expressed in free-form text, with the goal of incorporating these histories into LLMs to enhance recommendation performance. The UEM transforms user histories into concise representative embeddings, acting as soft prompts to guide the LLM. This approach addresses the computational complexity associated with directly concatenating lengthy histories.\\
\textbf{POD}: Chen~\etal~\cite{chen_KD} introduce PrOmpt Distillation (POD), is proposed to enhance the efficiency of training models. Experimental results on three real-world datasets show the effectiveness of POD in both sequential and top-N recommendation tasks.\\
\textbf{M6-REC}: Cui~\etal~\cite{cui2022m6rec} proposes M6-Rec, an efficient model for sample-efficient open-domain recommendation, integrating all subtasks within a recommender system. It excels in prompt tuning, maintaining parameter efficiency. Real-world deployment insights include strategies like late interaction, parameter sharing, pruning, and early-exiting. M6-Rec excels in zero/few-shot learning across tasks like retrieval, ranking, personalized product design, and conversational recommendation. Notably, it is deployed on both cloud servers and edge devices.\\
\textbf{PBNR}: Li~\etal~\cite{li2023pbnr} introduce PBNR, as a distinctive news recommendation approach using personalized prompts to predict user article preferences, accommodating variable user history lengths during training. The study applies prompt learning, treating news recommendation as a text-to-text language task. PBNR integrates language generation and ranking loss for enhanced performance, aiming to personalize news recommendations, improve user experiences, and contribute to human-computer interaction and interpretability.\\
\textbf{LLM-REC}: Lyu~\etal~\cite{lyu2023llmrec} introduce LLM-Rec, a method with four prompting strategies: basic, recommendation-driven, engagement-guided, and a combination of recommendation-driven and engagement-guided prompting. Empirical experiments demonstrate that integrating augmented input text from LLM significantly enhances recommendation performance, emphasizing the importance of diverse prompts and input augmentation techniques for improved recommendation capabilities with LLMs.\\
\textbf{Recommender Systems in the Era of LLMs}: Fan~\etal~\cite{fan2023recommender} survey the innovative use of prompting in tailoring LLMs for specific downstream tasks, focusing on recommendation systems. They review methods involving task-specific prompts to guide LLMs, aligning downstream tasks with language generation during pre-training. The study discusses techniques like ICL and CoT in the context of recommendation tasks within RecSys. It also covers prompt tuning and instruction tuning, incorporating prompt tokens into LLMs and updating them based on task-specific recommendation datasets.

\section{Fine-tuned LLMs in Recommender Systems}
\label{subsec:finetuned LLMs}
This section delves into the application of fine-tuned LMs in recommendation systems, exploring their effectiveness in tailoring recommendations for enhanced user satisfaction.\\
\textbf{TALLRec}: Bao~\etal~\cite{Bao} introduce a Tuning framework for Aligning LLMs with Recommendations (TALLRec) to optimize LLMs using recommendation data. This framework combines instruction tuning and recommendation tuning, enhancing the overall model effectiveness. Initially, a set of instructions is crafted, specifying task input and output. The authors implement two tuning stages: instruct-tuning focuses on generalization using self-instruct data from Stanford Alpaca, while rec-tuning structures limited user interactions into Rec Instructions.\\
\textbf{Flan-T5}: Kang~\etal~\cite{kang2023llms} conduct a study to assess various LLMs with parameter sizes ranging from 250 million to 540 billion on tasks related to predicting user ratings. The evaluation encompassed zero-shot, few-shot, and fine-tuning scenarios. For fine-tuning experiments, the researchers utilized Flan-T5-Base (250 million parameters) and Flan-U-PaLM (540 billion parameters). In zero-shot and few-shot experiments, GPT-3 models from OpenAI and the 540 billion parameters Flan-U-PaLM were employed. The experiments revealed that the zero-shot performance of LLMs significantly lags behind traditional recommender models. \\
\textbf{InstructRec}: Zhang~\etal~\cite{zhang2023recommendation} introduce InstructRec for LLM-based recommender systems, framing the task as instruction following. Using a versatile instruction format with 39 templates, generated fine-grained instructions cover user preferences, intentions, task forms, and contextual information. The 3-billion-parameter Flan-T5-XL model is tuned for efficiency, with the LLM serving as a reranker during inference. Selected instruction templates, along with operations like concatenation and persona shift, guide the ranking of the candidate item set.\\
\textbf{RecLLM}: Friedman~\etal~\cite{friedman2023leveraging} propose a roadmap for a large-scale Conversational Recommender System (CRS) using LLM technology. The CRS allows users to control recommendations through real-time dialogues, refining suggestions based on direct feedback. To overcome the lack of conversational datasets, the authors use a controllable LLM-based user simulator for synthetic conversations. LLMs interpret user profiles as natural language for personalized sessions. The retrieval approach involves a dual-encoder architecture with ranking LLMs explaining item selection. The models undergo fine-tuning with recommendation data, resulting in a successful proof-of-concept CRS using LaMDA for recommendations from public YouTube videos.\\
\textbf{DEALRec}: Lin~\etal~\cite{lin2024dataefficient} introduce DEALRec, an innovative data pruning approach for efficient fine-tuning of LLMs in recommendation tasks. DEALRec identifies representative samples from extensive datasets, facilitating swift LLM adaptation to new items and user behaviors while minimizing training costs. The method combines influence and effort scores to select influential yet manageable samples, significantly reducing fine-tuning time and resource requirements. Experimental results demonstrate LLMs outperforming full data fine-tuning with just 2\% of the data.\\
\textbf{INTERS}: Zhu~\etal~\cite{zhu2024inters} proposes INTERS as an instruction tuning dataset designed to enhance the capabilities of LLMs in information retrieval tasks. Spanning 21 search-related tasks distributed across three key categories—query understanding, document understanding, and query-document relationship understanding—INTERS integrates data from 43 distinct datasets, including manually crafted templates and task descriptions.INTERS significantly boosts the performance of various publicly available LLMs in information retrieval tasks, addressing both in-domain and out-of-domain scenarios.

\section{Evaluating LLMs in Recommender System} 
\label{section:evaluation}
This section assesses the performance of LLMs in the context of recommendation systems, evaluating their effectiveness and impact on recommendation quality.\\
\textbf{iEvaLM}: Wang~\etal~\cite{wang2023rethinking} introduce an interactive evaluation approach called iEvaLM, centered on LLMs and utilizing LLM-based user simulators. This method enables diverse simulation of system-user interaction scenarios. The study emphasizes evaluating explainability, with ChatGPT demonstrating compelling generation of recommendations. This research enhances understanding of LLMs' potential in CRSs and presents a more adaptable evaluation approach for future studies involving LLM-based CRSs.\\
 \textbf{FaiRLLM}: Zhang~\etal~\cite{Zhang_2023} posit that there is a need to assess RecLLM's fairness. The evaluation focuses on various user-side sensitive attributes, introducing a novel benchmark called Fairness of Recommendation via LLM (FaiRLLM). This benchmark includes well-designed metrics and a dataset with eight sensitive attributes across music and movie recommendation scenarios. Using FaiRLLM, an evaluation of ChatGPT reveals persistent unfairness to specific sensitive attributes during recommendation generation.\\
 \textbf{RankGPT}: Sun~\etal~\cite{sun2023chatgpt} proposed a instructional techniques tailored for LLMs in the context of passage re-ranking tasks. They present a pioneering permutation generation approach. The evaluation process involves a comprehensive assessment of ChatGPT and GPT-4 across various passage re-ranking benchmarks, including a newly introduced NovelEval test set. Additionally, the researchers propose a distillation approach for acquiring specialized models utilizing permutations generated by ChatGPT.


\section{Conclusion}
In summary, this paper explores the transformative impact of LLMs in recommender systems, highlighting their versatility and cohesive approach compared to traditional methods. LLMs enhance transparency and interactivity, reshaping user experiences dynamically. The study delves into fine-tuning LLMs for optimized personalized content suggestions and addresses evaluating LLM models in recommendations, providing guidance for researchers and practitioners. Lastly, it touches upon the intricate ranking process in LLM-driven recommendation systems, offering valuable insights for designers and developers aiming to leverage LLMs effectively. This comprehensive exploration not only underscores their current impact but also lays the groundwork for future advancements in recommender systems.

\bibliographystyle{ACM-Reference-Format}
\bibliography{main}

\appendix









\end{document}